\documentclass[twocolumn,showpacs,amsmath,graphicx,amssymb]{revtex4}

\usepackage{amsfonts}
\usepackage{amssymb}
\usepackage{amsmath}
\usepackage{graphicx}

\begin{document}

\title{Graphene physics and insulator-metal transition in compressed hydrogen}
\author{Ivan I. Naumov, R. E. Cohen  and Russell J. Hemley}
\affiliation{Geophysical Laboratory, Carnegie Institution of Washington, 5251 Broad Branch Road, Washington, D.C. 20015, USA}
\date{\today}

\begin{abstract}

Compressed hydrogen passes through a series of layered structures in 
which the layers can be
viewed as distorted graphene sheets. The electronic structures of these 
layered structures can be
understood by studying simple model systems-  an ideal single hydrogen 
graphene sheet and three-dimensional
model lattices consisting of such sheets. The energetically stable 
structures result from structural
distortions of model graphene-based systems due to electronic 
instabilities towards Peierls
or other distortions associated with the opening of a band gap. Two factors
play crucial roles in the metallization of compressed hydrogen: (i) 
crossing of conduction and valence bands in hexagonal or graphene-like 
layers due to topology and (ii) formation of bonding states with $2p_z$ $\pi$
  character.

\end{abstract}

 \maketitle
\section{Introduction}
The creation and characterization of metallic hydrogen under pressure has been described by  Ginzburg as one of the 
\textquotedblleft Key  Problems of Physics and Astrophysics" \cite{ginzburg}.
  Hydrogen in a metallic state is expected  to exhibit  high-T$_c$ superconductivity    \cite{ashcroft}
and other exotic properties, and achieving such a state is  thus of great fundamental interest.
At low temperatures hydrogen forms a simple hexagonal close packed structure with freely rotating molecules, called phase I \cite{mao}.
 At such temperatures,  the material transforms to the quantum broken symmetry phase (phase II) at pressures of 125 GPa
 and phase III at 150 GPa \cite{mao}.  The transition to phase III is characterized  by a strong change in the infrared  vibron absorption.

Recently, solid hydrogen has been intensively investigated in a new pressure-temperature
domain (200-360 GPa,  and just above 300 K), both experimentally
\cite{eremets,howie,zha, goncharov,zha2,loubeyre} and theoretically \cite{pickard1,pickard2,lu,labet1,labet2,lu2}.
 A new phase (phase IV) has been discovered above 220 GPa in the higher temperature regim  \cite{eremets,howie,zha2}.
 Though several different structures have been proposed for phase IV from first-principles calculations,
 \cite{pickard2,lu,lu2} there is agreement
 that  this phase can be viewed as a
 mixed layer structure, where  layers of weakly interacting  H$_2$ molecules
 are sandwiched between  graphene-like layers.

Since the band structures associated with $1s$  electrons in
 a hydrogenic honeycomb lattice and $2p_z$  electrons in
graphene are  symmetrically and topologically identical \cite{saito},
   two seemingly unrelated  topics in modern
condensed matter physics-- graphene physics and 
 hydrogen metallization are  intimately
interconnected. This interconnection is  wider than might appear at first sight.
In fact, the stability of solid hydrogen in structures consisting of  graphene-like hexagonal  sheets was predicted in 1981 \cite{lesar},
 well before the discovery of graphene. Using ab-initio calculations we show that  the electronic properties of \emph{all}  the proposed candidate structures for solid molecular hydrogen  at high density can be understood by studying a  single  graphene layer and/or simple systems composed of such layers. 
The structures can be viewed as distortions of underlying model graphene layers due to intrinsic electronic instabilities leading to the appearance of an excitonic insulator, Peierls distortion or molecular tilting.

The analysys leads us to identify two  new factors that can control  the metallization of compressed hydrogen. 
We show that  hydrogenic  hexagonal-  or graphene-like layers form band states with
Dirac-type cones where the bonding valence   and antibonding conduction bands inevitably touch each other
leading  to the zero-gap semiconductor or semi metallic  behavior.  This effect  explains  why  many proposed structures with  nearly the
same energy and symmetry exhibits nevertheless distinct band structures in the vicinity of the Fermi level E$_F$.
We also show that   lowering of the bonding  $2p_z$  and other states  comes into play above  200 GPa
when the energy difference between the 1$s$ and 2$p$ atomic orbitals  become
comparable with the bandwidths of  the 1$\sigma_{g}$  and 1$\sigma_{u}^*$ states. The calculations provide new insights into pressure-induced 
metallization and the interpretation of recent experimantal results.
\section{Methods}
Calculations were performed using density functional theory implemented in the ABINIT
package.  A 16$\times$16$\times$1 Monkhorst-Pack $\boldsymbol{k}$-point grid has been used in the case of ideal  hydrogenic
honeycomb lattice containing only two atoms per cell;
  approximately the same $\boldsymbol{k}$-point
density was kept in going from the 2D honeycomb lattice to 3D systems. The sheets have been
simulated by a slab-supercell approach with the inter-planar distances of 25 a$_B$ to ensure negligible
wavefunction overlap between the replica sheets. For the plane-wave expansion of the valence
 and conduction band wave-functions, a
cutoff energy was chosen to be 40 Ha. A
 norm-conserving pseudopotential with a
cutoff radius of  0.5  a.u. was first generated using OPIUM codes and then
 used, along with the Wu-Cohen exchange
and correlation functional  \cite{wu}. The obtained results were tested against all-electron FLAPW calculations,
and  very good consistency between  the two was found.

\section{Results}
\subsection{A single graphene layer}

We begin with a discussion of the  electronic and atomic properties of a  2D  honeycomb hydrogenic lattice under pressure
which we mimic by changing the nearest distance between two hydrogen atoms $b$.
 At zero pressure, this distance  was found to be  1.177 \AA{},
which is considerably larger   than that found for a 6-membered hexagon with D$_{6h}$ symmetry
 (0.992 \AA{}) \cite{labet2}. This  reflects the fact that chemical bonding in an isolated hexagon
is stronger than in a periodic network of such hexagons, where each atom must share its  electron with two neighbors.

\begin{figure}[h]
%\vspace{1 cm}
\begin{centering}
\includegraphics
[width=10.0cm]{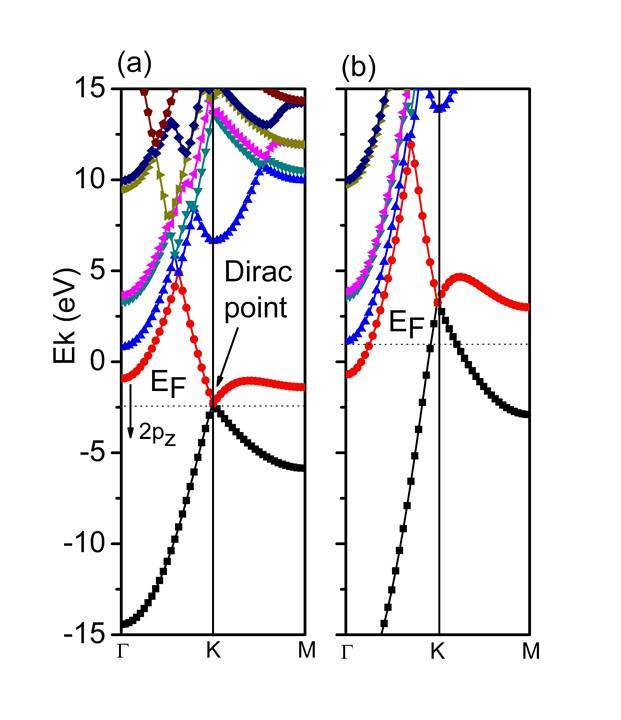} \caption{Calculated band structure of H-graphene  for two different bond lengths:  (a)  $b$=1.23  \AA{}  and   (b)  $b$=0.94
\AA{}. The vertical arrow in (a) indicates the bonding conduction band $2p_{z}$
 moving quickly downward as the lattice parameter decreases.
At $b$=1.10 \AA{}   this band passes the Dirac level becoming lower than it, as seen in (b). The equilibrium $b$=1.177  \AA{}. }
 \label{fig:fig1}
%\vspace{0.05 cm}
\end{centering}
\end{figure}

The band structure of an ideal 2D hydrogenic honeycomb lattice  is shown in  Fig.~\ref{fig:fig1}.
 When the bond length $b$ is larger than  1.10  \AA{},  this structure
is similar to  that in carbon graphene, where  the conduction and valence  bands  touch each other in a linear fashion at two inequivalent Dirac points, $\mathbf K $ and $\mathbf K'$,
connected by  time-reversal  symmetry (Fig.~\ref{fig:fig1}a). However,
 the next (and higher)  energy level(s) at $\Gamma$   are  very sensitive to the lattice spacing and quickly move downwards  as the  lattice parameter decreases,
as  shown by a vertical arrow in Fig.~\ref{fig:fig1}a.   As a result, at $b$ =1.10 \AA{},  the second level  passes the Dirac level becoming lower than it,  and the system undergoes a transition from a zero-gap semiconductor to a semimetallic
state,   Fig.~\ref{fig:fig1}b. This level is the bottom of the bonding 2$p_z$ band, so that the
 transition is accompanied by the  charge  flow  from the antibonding 1$s$ orbitals to the  bonding   2$p_z$ orbitals.
\begin{figure}[h]
%\vspace{1 cm}
\begin{centering}
\includegraphics
[width=9cm]{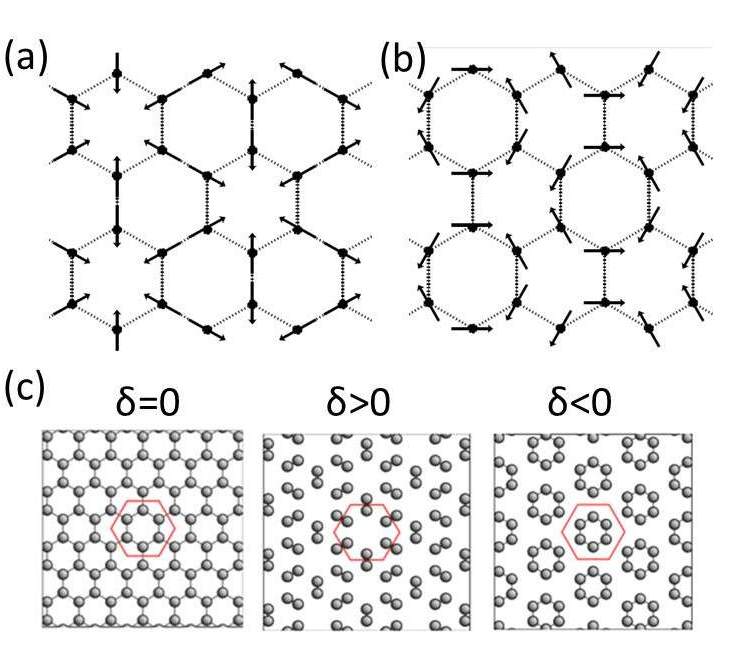} \caption{  Peierls  distortions of a honeycomb lattice, leading to   a
$\sqrt{3}
 \times \sqrt{3}$ superstructure and described by  an amplitude $\delta$.
(a) TO- distortion, (b) LA- distortion; modified after Ref. \cite{gunlycke}. (c) TO-distortions for different $\delta$;  $\delta$=0 corresponds to a
pristine
 honeycomb lattice  where a non-primitive unit cell
is chosen in the form of hexagon. In the case of $\delta>0$  the initial lattice dimerizes, so that the initial
atomic hexagon  expands.  In the case of $\delta<0$,
 the initial lattice hexamerizes, and  the initial atomic  hexagon  decreases in size.  Modified after Ref. \cite{lee}.}
\label{fig:fig2}
%\vspace{0.05 cm}
\end{centering}
\end{figure}
\begin{figure}[h]
%\vspace{1 cm}
\begin{centering}
\includegraphics
[width=9 cm]{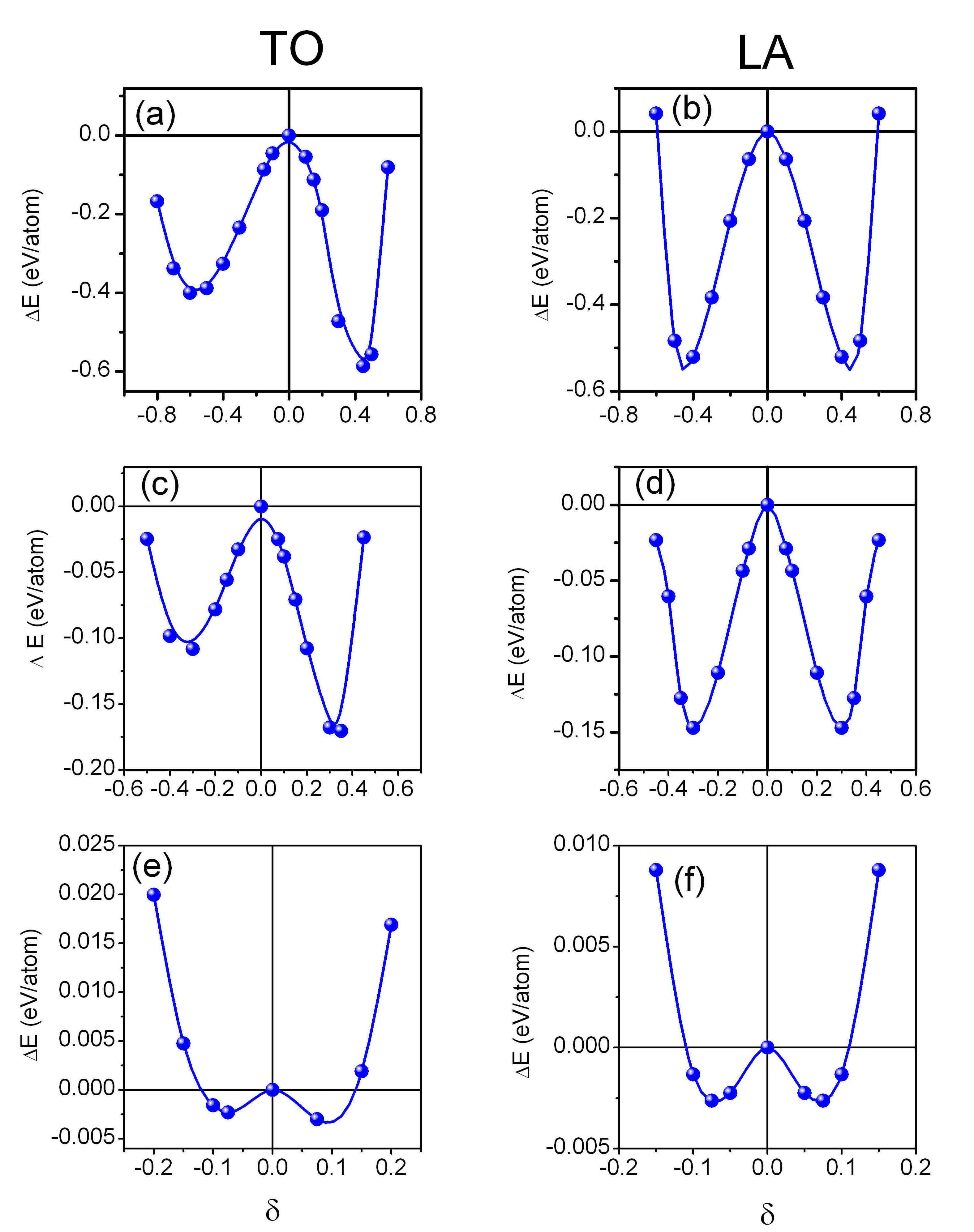} \caption{ Total energy vs  TO  and LA  mode amplitude $\delta$  for
different  initial bond lengths   $b$: (a,b) 1.40  \AA{},  (c,d) 1.177  \AA{}  and  (e,f)  1.0 \AA{}.  The  TO  and LA
modes are actually symmetrized to produce the distortions of  the  D$_{6h}$ and D$_{3h}$ symmetries,
respectively.}
 \label{fig:fig3}
%\vspace{0.05 cm}
\end{centering}
\end{figure}
\begin{figure}[h]
%\vspace{1 cm}
\begin{centering}
\includegraphics
[width=9cm]{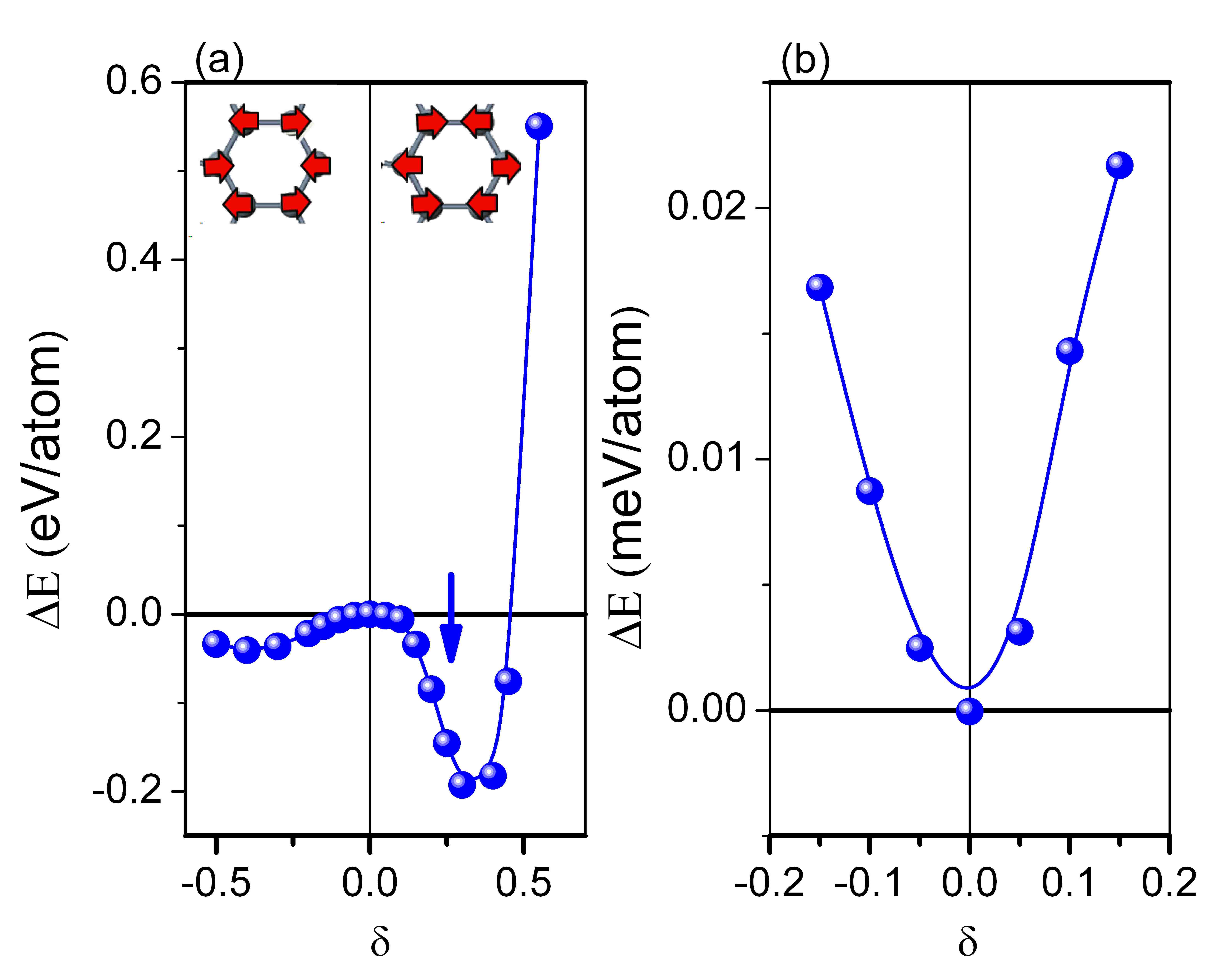} \caption{Total energy vs   $\Gamma$ optical phonon amplitude $\delta$ for
different  initial bond lengths $b$: (a)  1.177 \AA{}   and  (b)  1.0 \AA{}. The inserts illustrate that in  case of
$\delta>0$  two  nearest hydrogen atoms in each unit cell   approach  each other, whereas for  $\delta<0$  they
move away  from each other. The vertical blue arrow indicates the moment when two Dirac points
merge to open a band gap.}
 \label{fig:fig4}
%\vspace{0.05 cm}
\end{centering}
\end{figure}

The  ideal honeycomb lattice is   unstable with respect to    structural distortions. The existence of
Dirac cones implies that there is   nesting between  the  valence and conduction  bands, which can be expressed as
\begin{equation}
-\epsilon_{i} (\boldsymbol{k})=\epsilon_{j} (\boldsymbol{k+K-K^{\prime}}),
 \label{eq:1}
\end{equation}
where the valence ($i$) and conduction ($j$) band energies are measured relative to  E$_F$; note that the difference $\boldsymbol{K-K^{\prime}}$ is again a Dirac point plus a reciprocal lattice vector.
In the case of carbon graphene, the condition  (\ref{eq:1}) leads to   strong electron-phonon coupling and  the $\boldsymbol K$ point  phonon branch softening
(Kohn anomaly) \cite{mafra}.  Such an anomaly should also exist in the  hydrogen lattice and even be stronger  in the absence of $sp^2$ bonding.  Thus,
we  expect the lattice  to be  unstable against  2D Peierls-like    distortions \cite{lee} which are  the  frozen symmetrized combinations
of the  soft TO and LA phonon modes at $\boldsymbol{K}$-points \cite{gunlycke}; they break the  translation symmetry and lead to a
  $\sqrt{3}\times \sqrt{3}$ superstructure.

We now  consider the stability of the  lattice against such distortions  for different lattice parameters.
The TO (or so-called Kekul\'{e}) distortion    can be introduced  via   in-plane atomic displacements preserving
the
 initial D$_{6h}$  point symmetry. It can be described  by only one structural parameter or amplitude, $\delta$,  which can be both positive and negative
 (Fig.~\ref{fig:fig2}). Positive  $\delta$'s  lead to the dimerization of a initial  atomic lattice where the intramolecular distance is $ b(1-\delta)$, and
the intermolecular distance  $ b(1+\delta/2)$, where  $ b$ is the initial bond length.  In contrast, negative $\delta$'s  describe the process of breaking of the
initial lattice into smaller hexagons (\textquotedblleft hexamerization") where the nearest distance between  different  hexagons is
 $b (1-\delta)$,  whereas the  bond length   inside new and smaller hexagons is given by $b (1+\delta/2)$.

Similar to the TO Peierls distortions, the LA counterparts can also be described by a single amplitude $\delta$
(Fig.~\ref{fig:fig2}).  The LA distortions reduce the symmetry from D$_{6h}$ to D$_{3h}$,  and the corresponding
energy profile is symmetric with respect to the change of the sign of $\delta$. Both  $\delta >$ 0 and   $\delta
<$ 0 lead to the dimerization of hydrogen atoms where the resulting intramolecular and intermolecular distances
are  $ b(1-|\delta|)$ and $ b(1+|\delta|)$, respectively.

The energy profiles calculated as a function of the amplitude $\delta$  for different initial bond lengths $b$  are shown in  Fig.~\ref{fig:fig3}.
In the case of TO distortions, such profiles  exhibit two asymmetric minima, one  for
 $\delta >$0 and the other for  $\delta <$0. The first one corresponding to  dimerization of hydrogen
atoms is more stable than the second leading to separated hexagons. This is in agreement with early results
\cite{lesar} showing that the energy of the  hexagonal H$_6$  complex is higher than that  of three molecules
H$_2$. The minimum for $\delta >$ 0 is not only deeper but also sharper since  here  the atoms in each
 pair  move towards each other and some point start mutually  repelling  as the separation becomes  somewhat less than 0.75  \AA, the bond length in an isolated   H$_2$.

The    profiles  for the LA Peierls distortions are similar
to that for TO counterparts with $\delta >$0; this is not surprising  because  in both
cases they describe the process of dimerization. Again,
the minima correspond to a intramolecular distance of
 0.75 \AA{}    and  are sharp in form due to the repulsion  of the
H atoms below this critical separation.  Figure ~\ref{fig:fig3} shows that
the wells become progressively shallower and closer to each other as the initial bond length  is decreased. Just below  $b=$ 1 \AA{},  the minima
 merge and the  energy profile  becomes single-welled.  This point correlates well with the moment when  the  bonding $2p_{z}$-band  at
$\Gamma$ passes the Dirac energy (1.1.\AA{}).   This fact can be easily understood. The Peierls distortions
 mix the unperturbed $\boldsymbol{K}$ and $\boldsymbol{K^{\prime }}$ Bloch states and break the 4-fold degeneracy at $\Gamma$ of a  pristine
 honeycomb lattice into two 2-fold generacies opening the gap and lowering the total energy.  But the latter can happen
  only when the Dirac point coincides with the Fermi level.
   The is not the case when
the $2p_{z}$ band near  $\Gamma$ becomes  noticeably lower than the Fermi level, thus eliminating   the driving force for the
Peierls distortions. Though freezing of these modes opens up  a gap at Dirac points and  lowers the total energy,
it does not  effect much the pace of  the lowering of the $2p_{z}$ states.

Analysis shows that the graphene lattice can be  unstable against other distortions leading to the dimerization of
hydrogen atoms. These distortions, however, become  energetically  less favorable than the  Peierls ones 
for   the initial bond lengths $b$ shorter than $\sim$ 1.15  \AA{}. 
To illustrate this, consider, for example,  the  simplest path for the  association of hydrogen atoms into
molecules  when  two  nearest  hydrogen atoms in each unit cell  move  towards each other. The distortion can be visualized
as a freezing of the optical phonon mode at the $\Gamma$ point. As for   the TO($\boldsymbol{K}$) and
LA($\boldsymbol{K}$) symmetrized modes, this distortion  can be described by a single parameter  $\delta>0$
defining the distance  between the  initially nearest atoms, $b(1-\delta)$ (see inset in Fig.~\ref{fig:fig4}a).  
For comparison, we will also
consider  the case $\delta<0$  when the atoms  move apart  so that the  initial hexagonal lattice transforms
toward a rectangular lattice corresponding to $\delta=-0.5$.
 The energy as a function of  $\delta $  is shown in Fig.~\ref{fig:fig4}; it is 
similar to that for TO($\boldsymbol{K}$) distortions.  At $b$=1.177  \AA{}   it exhibits two
highly assymmetric minima, for positive and negative $\delta$. The  first  one associated with the pairing of the
hydrogen atoms is significantly deeper. It is  even slightly deeper than the  minima for the Peierls
distortions  corresponding to the same initial  $b$  (Fig.~\ref{fig:fig3}c,d). However,   as  seen from Fig.~\ref{fig:fig4}b, for $b$=1.0 \AA{},    
double well profile is already turned into a single-welled,  in contrast with the profiles in Fig.
\ref{fig:fig3}e,f.  This proves  that the Peierls distortions energetically are the most
preferable distortions for the relartively short initial bond lengths $b$.
 Note that  in the case of  $\delta<0$ for  $b$=1.177  \AA{}  the minimum   forms   just before  the structure
becomes rectangular. This minimum is very shallow because the  intramolecular distance on this path can not be
smaller than $b\sqrt{3}/2 \sim$ 1.02 \AA{} and the molecules with the  optimal bond length $\sim$  0.75 \AA{}
can not form.

A closer examination of the  Fig.~\ref{fig:fig4} reveals that the position of the minimum in  the panel (a)  for $\delta>0$ is
 very close to the bond length in an isolated  molecule H$_2$, 0.75 \AA{}. Before this minimum is reached, the
energy curve exhibits  the steepest   fall  exactly at  the moment  when  two Dirac points shifted from their
initial positions  merge at the point $(\boldsymbol{G}_{1}-\boldsymbol{G}_{2})/2$, where $\boldsymbol{G}_{1}$ and
$\boldsymbol{G}_{2}$ are  two shortest reciprocal lattice vectors  forming  the angles 60$^0$ and --60$^0$  with
the line connecting two nearest atoms separated by the distance  $b(1-\delta)$. This suggests that an opening a
gap belongs  to  the major stabilizing factors in compressed hydrogen.
\begin{figure}[h]
%\vspace{1 cm}
\begin{centering}
\includegraphics
[width=9cm]{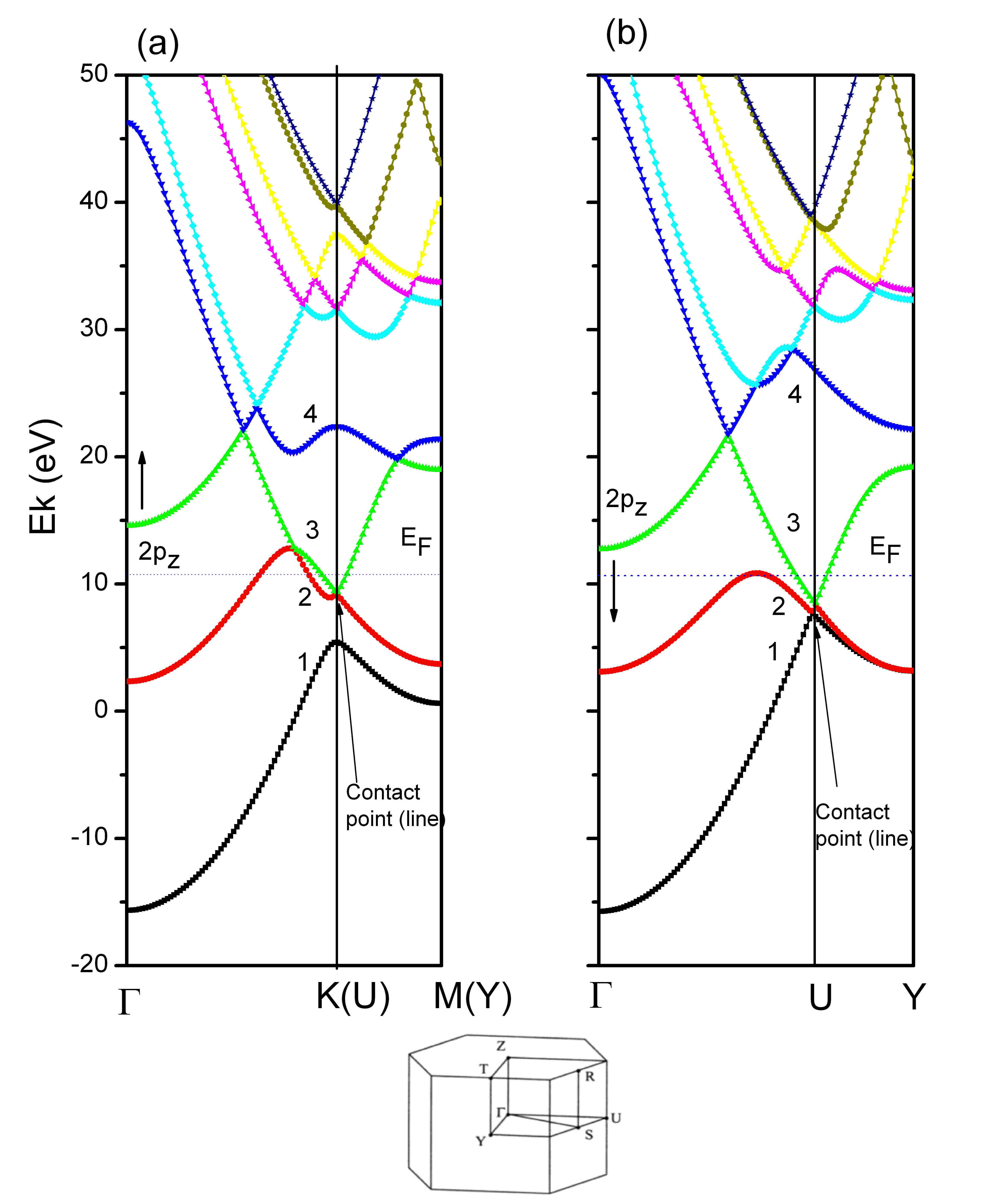} \caption{Calculated band structures at 260 GPa:  (a) for  P6$_3$/mmc 
 and  (b) for
  Cmca-4. For the sake of comparison, the high symmetry points in the hexagonal Brillouin zone points of  P6$_3$/mmc structure
 are indicated by double labels with  the  orthorhombic labels in parentheses.}\label{fig:fig5}
%\vspace{0.05 cm}
\end{centering}
\end{figure}

We point out  that  the  states  with Peierls distortions can be equally interpreted as the excitonic
insulators characterized by charge-density oscillations,  as discussed
 by  Halperin and Rice \cite{halperin}. Indeed, the condition   (\ref{eq:1}) also means that the electron and hole Fermi surfaces are identical in size and shape--a necessary condition for the formation of the
excitonic insulators. Another necessary  condition--weak   screening of Coulomb interactions between electrons and holes--
 is also fulfilled in hydrogen graphene due  the vanishing density of states at the Dirac points and the single-atom thickness of the system.  Therefore,  excitonic states are  likely to form   in compressed
hydrogen, especially at low temperatures.

The (Peierls) distortions  considered above are not the only ones capable of inducing 
an energy gap at the Fermi level in 2D honeycomb hydrogenic lattice via $\boldsymbol{K}$ and
$\boldsymbol{K^{\prime }}$ mixing.  Other  distortions with non-zero wavevectors can also do this, provided that
their amplitude exceeds some (usually small) critical value  \cite{naumov}. Among such Peierls-like distortions
are the $3 \times  2$ superstructures where the length of the super-period along a zigzag type directions larger
than that in honeycomb lattice by a factor of 3,  whereas the super-period along an armchair type direction by a
factor of 2.

\subsection{Stacked graphene layers }

 We now consider 2D   hydrogen honeycomb lattices stacked in AB (Bernal \cite{charlier}) fashion to produce a  3D graphite hexagonal structure with P6$_3$/mmc symmetry. In the 3D lattice there are four atoms per unit cell, twice as many in the honeycomb layer.  Each layer must give rise to two three-dimensional states near the $\boldsymbol{K}$  and $\boldsymbol{K'}$ points of the hexagonal Brillouin zone. So the  four initial single-layer  bands   mix  together by  interlayer interactions
 in order to form the band structure in the vicinity of  $\boldsymbol{K}$  and  $\boldsymbol{K'}$ points  \cite{slonczewski}.
 Though the AB  stacking destroys the sublattice or inversion symmetry in each single layer, the necessary condition  for the formation of Dirac degeneracy points,  such points
 nevertheless do survive in the  3D structure    due to the interlayer interactions   \cite{slonczewski},  as it is seen from Fig.~\ref{fig:fig5}a.
 Moreover,  the degeneracy points merge together along the vertical edges HKH  of the   hexagonal Brillouin zone, thus forming      band-contact lines.
These lines exhibit  the following interesting topological property: the line integral of the Berry connection for any curve enclosing them is $\pm \pi$   \cite{miktik}.

In  Fig.~\ref{fig:fig5}a we present  the band structure for P6$_3$/mmc calculated at a pressure of 260 GPa. We
indicate the energy bands in the vicinity  $\boldsymbol{K}$  and  $\boldsymbol{K'}$ points by by 1, 2, 3 and 4 in
order to stress that they are  originated  from the mixing the initial layer states.  It is remarkable that their
forms are in  qualitative agreement with  that predicted  in Ref.  \cite{slonczewski}  for carbon  graphite  on
the basis of tight-binding calculations. From  Fig.~\ref{fig:fig5}a  it is clearly seen that at $\boldsymbol{K}$,
the 2-fold degenerate state lies  between the two single levels, thus  forcing  the valence and conduction bands
to touch \cite{slonczewski}. Since  the band 2 exhibits a local maximum near the $\boldsymbol{K}$  point, the
Fermi level is forced also to cross the band 3 as well. As a result, in contrast to the 2D hydrogen honeycomb
lattice, the system represents a semimetal without  involving of  the  2$p_z$  states.

 The  bonding  $2p_z$ states  in  the P6$_3$/mmc  structure  fall at  the bottom of the third band  near  $\Gamma$,
 Fig.~\ref{fig:fig5}a. At   P=0, their energies practically coincide with the Fermi level E$_F$ but gradually move away from the
 latter as the lattice parameter is decreased. As a result, at  260 GPa,   they become  well above the  E$_F$, by 3 eV.
  This is not the case, for example, in the  Cmca-4 phase, as we will see below.
We stress  that the degenerate states on the  contact-band lines (HKH) in the  P6$_3$/mmc structure
transform according to a  two dimensional representation and the  lines
 themselves coincide with  the 3-fold symmetry  lines. However, due to their topological properties, these  contact
  lines are stable against any (not too severe)  lattice distortions including those breaking 3-fold symmetry.
  As a result the  contact-band lines may  shift from the high-symmetry lines in the Brillouin zone becoming curvy
  in shape  like spirals  (such a spiral     ending  up on the faces of the Brillouin zone is found, for example, in
   the  rhombohedral graphite  with ABC stacking \cite{charlier, mcclure}).

\subsection{Analysis of candidate structures}

 Though a graphene motif is
most evident in the candidate structures  Cc, Pc and  Pbcn  for phase IV \cite{pickard2,lu,labet1,labet2}, it can be also found
  in other candidate structures proposed for dense
 hydrogen.  Thus,  the distorted   hexagons are clearly seen  in  the layered C2/c and Cmca-4  structures. The former is the   most plausible candidate
  for the low temperature phase III and the latter is a predicted at higher pressures and low temperatures \cite{pickard2,lu, goncharov}.
The intermolecular
 distance in graphene-like layeres of  high-pressure
structures  like C2/c, Cc, Pc and  Pbcn  is of the order 1.0--1.1 \AA{}     \cite{pickard2,lu,labet1,labet2}. This is 
noticeably shorter than the minimal distance separating  two hydrogen atoms from the neighboring layers (1.4
--1.5 \AA{}).  We expect  therefore that  such layers play a key role in  forming their electronic properties and atomic
structures.

\begin{figure}[h]
%\vspace{1 cm}
\begin{centering}
\includegraphics
[width=9cm]{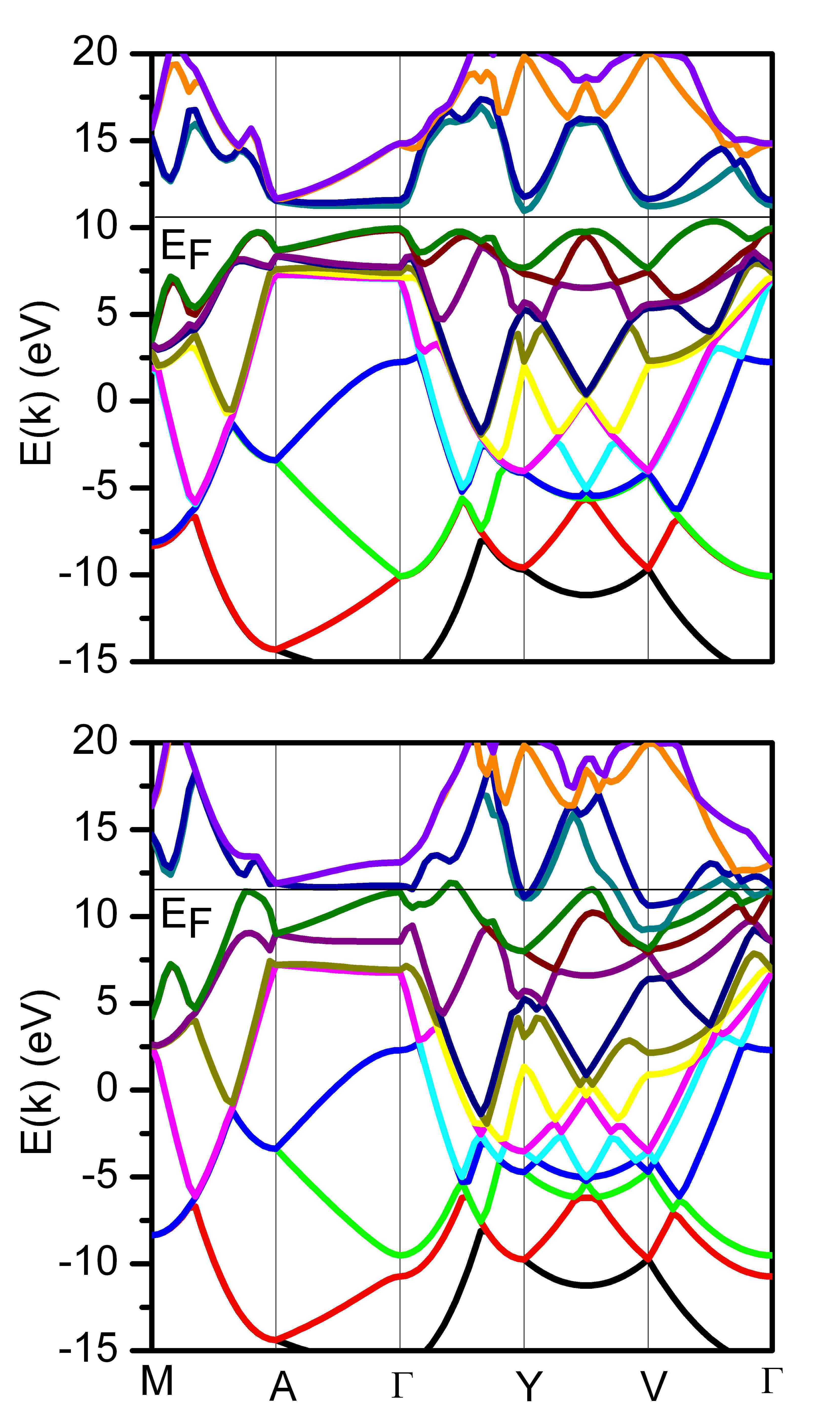} \caption{The band structures  of  the C2/c  phase at 300 GPa phase  (a)  and of
an artificial phase  composed from four ideal graphene layers ABCD in such a  way that the resulting symmetry is
also C2/c (b).  The artificial phase, like  C2/c, has 24 atoms per cell and its  lattice parameters are also
identical to that for C2/c structure. The notation of the symmetrical points is borrowed from the
Ref.\cite{pickard2}}
 \label{fig:fig6}
%\vspace{0.05 cm}
\end{centering}
\end{figure}
\begin{figure}[h]
%\vspace{1 cm}
\begin{centering}
\includegraphics
[width=9cm]{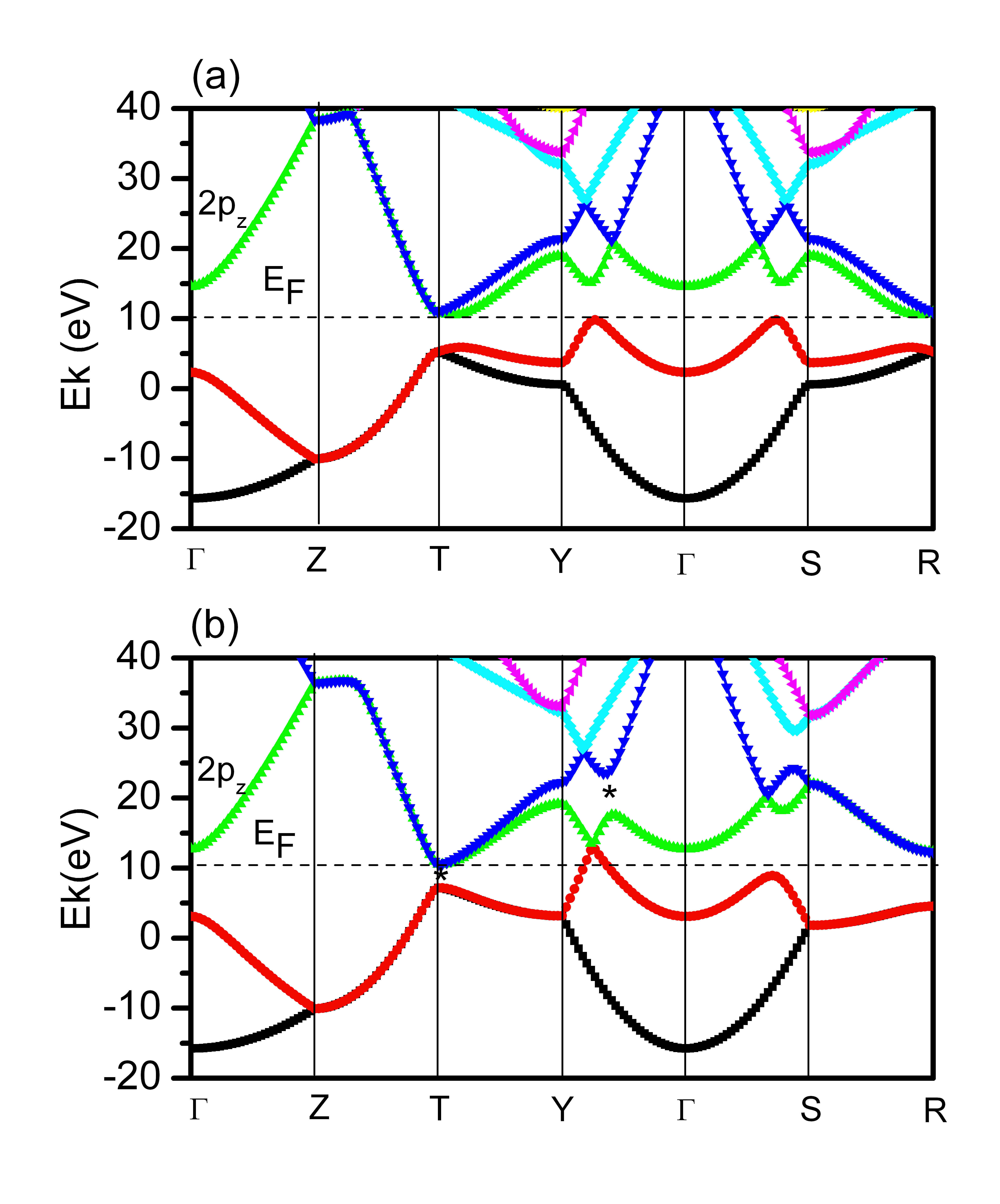} \caption{ Same as in Fig.~\ref{fig:fig5}, but for other directions in the
orthorhombic Brillouin zone. In  panel (b), the star * indicates the band gaps that appear due to molecular
tilting, $\theta \neq 0$. } \label{fig:fig7}
%\vspace{0.05 cm}
\end{centering}
\end{figure}
\begin{figure}[h]
%\vspace{1 cm}
\begin{centering}
\includegraphics
[width=9cm]{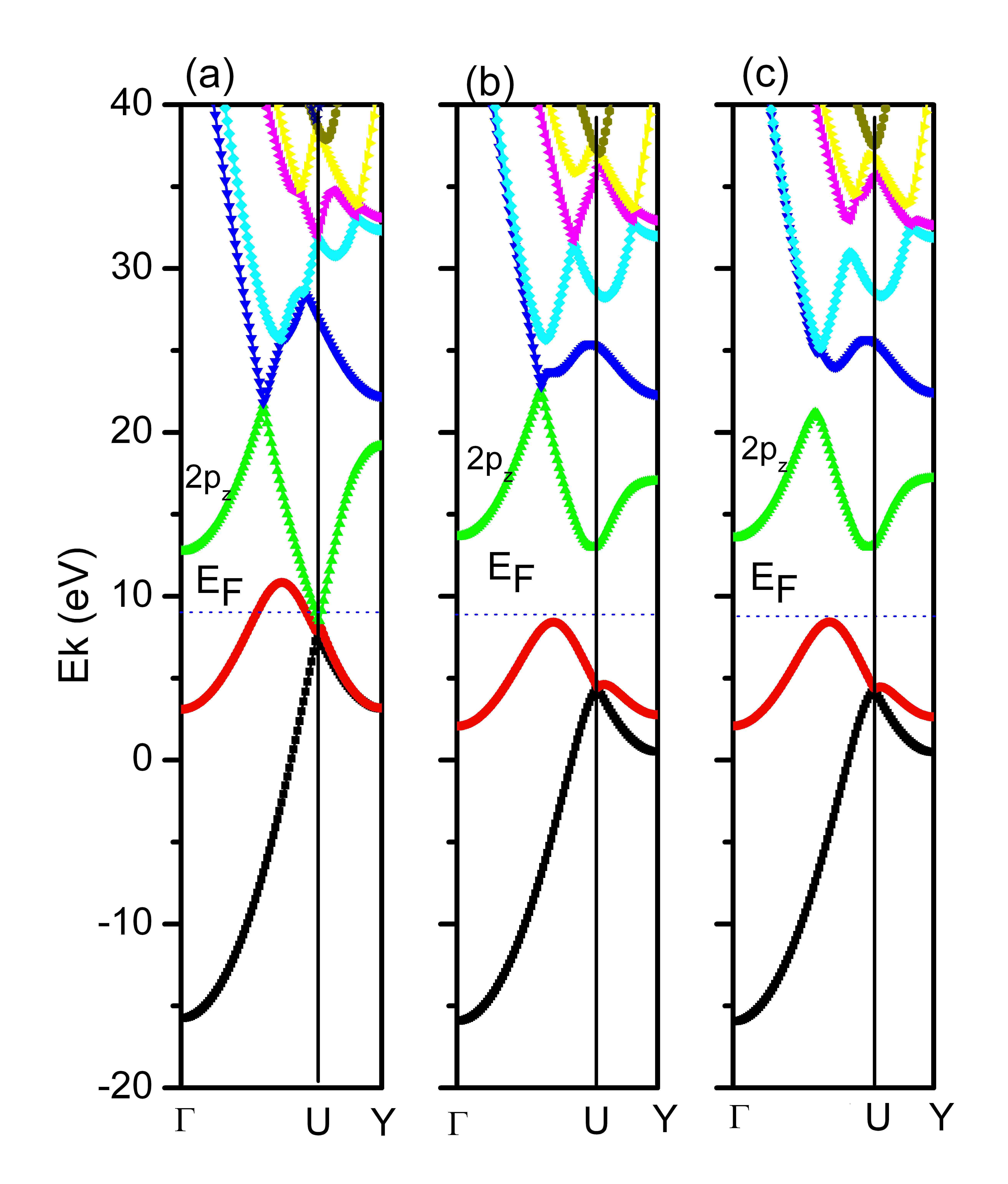} \caption{Calculated band structures:  (a)  for Cmca-4 at P=260 GPa,   (b)
for  Cmc2$_1$,  and (c) for C2/m phases. In going from Cmca-4 to Cmc2$_1$ the lattice parameters, intramolecular
distances and tilting angles are kept the same. The only difference between these structures is that in the
latter  the centers of molecules lie almost on an ideal  hcp lattice. The C2/m structure differs from   Cmc2$_1$
only by mutual orientation of two molecules in the unit cell: whereas in  the Cmc2$_1$  the molecular axes are
tilted by the angles $\theta$ and -   $\theta$ from the $xy$  plane,   in the C2/m  the axes
  are parallel, i.e. molecules   tilted by the same angle  $\theta$, $\theta  \sim 30 ^0$.  }
 \label{fig:fig8}
%\vspace{0.05 cm}
\end{centering}
\end{figure}

Among these  candidate structures,  C2/c  phase  would appear to be a very large  distortion of an   ideal
honeycomb lattice. Yet such a comparison is possible. To mimic this phase we designed an artificial phase
where four ideal graphene layers stacked equidistantly in a ABCD fashion in such a way that resulting symmetry is
again C2/c.   In fact, this model structure can be continuously transformed within the given symmetry into  C2/c
 by moving the atoms inside the unit cell. As it seen from the Fig.~\ref{fig:fig6}, the band structure of the
model system does mimic well that of the C2/c, especially above and below the Fermi level. The main difference,
of course, is in the vicinity of the E$_F$ where the model structure exhibits, in contrast to C2/c, an
overlap of the conduction and valence bands being a metal. Such a difference is not surprising because each
layer in the C2/c structure can be considered as a $\sqrt{3}\times \sqrt{3}$ superstructure relative to an ideal
graphene sheet; in this superstructure  the initial graphene $\boldsymbol{K}$ and $\boldsymbol{K^{\prime }}$states
are mixed to form a band gap. Hence, according to Sec. A, this structure can be viewed as a Peierls  or excitonic
insulator phase.

We find that the band structures of all the other candidate phases   can be understood in this way, including
mixed phases, like Cc, Pc and Pbcn \cite{pickard2,lu,labet1,labet2}, where   the layers of unbound
H$_2$ molecules are sandwiched between the graphene-like layers.  The situation with Pbcn structure, for example,
is similar to that for  C2/c. Here again the insulating behavior of the systems can be easily explained: each layer
in Pbcn structure can be presented as a ${3}\times 2$ superlattice stabilized by Peierls-like distortions that mix
the Dirac states $\boldsymbol{K}$ and $\boldsymbol{K^{\prime}}$ and open an energy gap (see Sec. A).

Let us consider in more detail  the    Cmca-4 structure,   which    geometrically  is    close to the hydrogen graphite
structure P6$_3$/mmc discussed  in Sec. B.
 Like P6$_3$/mmc, it  can also  be considered as   a layered  structure   with ABAB stacking    where the H$_2$ molecules  form an angle $\theta \sim 30^0 $ with
respect to  the $xy$ plane. As seen from  Fig.~\ref{fig:fig5} (b)
this structure,  like P6$_3$/mmc,   exhibits  band-contact lines
which  now coincide well  with the vertical lines passing through the U points.   The character of splitting of energy levels at U point is
 different from that at the $\boldsymbol{K}$ and $\boldsymbol{K^{\prime }}$  points  in P6$_3$/mmc structure: the degeneracy point lies below the single levels.
  Another  difference is that now the $1s$ orbitals associated with  the  band  4 near the U point
  are strongly  hybrydized with bonding $2p_z$ orbitals. It is the hybridization
  that makes  the $2p_z$ orbitals to behave  differently  in the comparable systems as the pressure increases.
   We found that at 260 GPa the 2$p_z$  level at $\Gamma$ in Cmca-4 phase is by 4 eV above the  E$_F$. However,
   the former moves quickly  down with pressure and passes the   E$_F$ at $\sim$ 300 GPa. But even before this
   critical pressure the system is already in a semimetallic state due shifting of the E$_F$ relative to the
   contact point (Fig.~\ref{fig:fig5}b).

As analysis shows, the  above mentioned hybridization  is due to the orientational tilting of
  hydrogen molecules  enabling  mixing of $1s$ and $2p_z$ orbitals to form $sp$ like orbitals.
 This tilting  opens up a gap  at the Fermi level at the T point,  Fig.~\ref{fig:fig7}b. The rotation of hydrogen molecules
  tends to prevent the the system from being (semi)metallic. Despite this fact  the valence and conduction bands cross each other
  near the U points due to topological reasons, Fig.~\ref{fig:fig5}b .

This crossing, however, can be easily destroyed by sliding of alternate layers of towards the polar Cmc2$_1$
structure, Fig.~\ref{fig:fig8}b.  The reason is that such a sliding breaks an effective  in-plane  inversion
symmetry- the center of symmetry of a 2D lattice obtained by projection of two nearest layers  on the $xy$
plane.  The energy gap at U point also opens  up in the C2/m structure where such a symmetry center  is also
broken, Fig.~\ref{fig:fig8}c. Though  the opening of a gap due to a Cmca-4 to Cmc2$_1$ structural
transformation has already been discussed in the literature (e.g., Ref.\cite{neaton}), the  physical
reason for this was not clear.

\section{Discussion}
Our results suggest two possible  mechanisms of metallization of compressed hydrogen. The first one is   related
to the fact that in the systems with honeycomb-structured or  graphene-like layers a metallic electronic
structure occurs because of symmetry and topology,  as,  for example,  in graphite-structured hydrogen or  the
Cmca-4 phase. The first metallization mechanism involves only $1s$ bands, whereas the second  mechanism involves 1s
valence states and   $2p_z$  conduction states.   Under pressure, the bonding  $2\pi$ states associated with the
atomic  $2p_z$  orbitals become lower in energy than the  antibonding  1$\sigma^{\star}_u$   states originating
from the $1s$  orbitals.  The fact that the  $2p_z$  bands come into play at elevated pressures is not surprising
--the energy difference between the $1s$ and $2p$ atomic orbitals  becomes comparable with the widths of the
1$\sigma_g$  and 1$\sigma^{\star}_u$  bands  ($\sim$10 eV). The decrease in the  bandgap  due to these two effects
can (and usually does) widen again due to  Peierls-like distortions or molecular tilting;  this delays the
transition to higher pressures at low temperatures. 

Despite the similarities, there are fundamental differences between 
   H-graphene and  C-graphene. The two have  a similar electronic structure  and 
represent a zero-gap insulator with topologically protected Dirac points at $\boldsymbol{K}$ and
$\boldsymbol{K^{\prime }}$. However,  in C-graphene  the $sp^2$ states provide most of the structural stability
and completely suppress the Peierls instability associated solely with the   $2p_z$ states, in H-graphene the
relationship between the electronic and atomic structures is more subtle, because the  same $1s$-electrons provide
the bonding and structural stability. As a result,  ideal H-graphene sheets find a variety of different ways to
open a band  at the Dirac points  gap and reduce the total energy. The most effective distortions are the Peierls
(or Peierls-like) distortions   that induce the energy gap via $\boldsymbol{K}$ and $\boldsymbol{K^{\prime }}$
mixing. These displacements  lead to the formation of superlattices multiples of 3 or $\sqrt 3$ from the
primitive graphene cell; such superlattices indeed were found in the proposed phases for dense hydrogen(e.g.,
Pbcn, Cc, C2/c). The systems also avoid the crossing of 1s and  $2p_z$ bands  by mixing them, as in  the case of
diamond-type Si where the mixing of $3s$ and $3p$ bands form a gap along the $\Gamma$-X direction \cite{mcmahan}. To
enable the $1s-2p_z$ mixing in a graphene layer,  the  H$_2$ pairs should tilt out of the ideal honeycomb plane.
Such a tilting was found, for example, in the molecular layers of the Pbcn structure and in all layers of the C2/c
structure.

In summary, we have shown that compressed hydrogen can reach a semimetallic state via (i)  the formation of
bits of Fermi surface representing  the intersections of Dirac cones from graphene related structures (near the
$\boldsymbol{K}$-point in ideal graphene) or (ii) closing of an indirect band gap between  the  valence and
conduction bands originating principally  from $1s$ and  $2p_z$  atomic orbitals, respectively. A 
consideration of quantum and thermal fluctuations  should not change our conclusions. These fluctuations  will
serve to broaden the band states and initiate band overlap at lower compressions than indicated here, as shown in
\cite{morales}. However, the discontinuity in the potential at E$_F$
 or electron self-interaction will increase the gap. These two effects of opposite sign will at least partly cancel,
 as discussed in Ref. \cite{pickard1}. In any event, the  graphene-like  layers will dominate the behavior of
 high pressure hydrogen, its metallization, and other optical  and electronic properties.

\section{Acknowledgements}
We thank I. Mazin,  C. Zha and S. Mandal for helpful discussions. 
This research was supported by EFree, an Energy Frontier Research Center funded by the U.S. Department of Energy, Office of Science, Basic Energy Sciences under Award DE-SC0001057.  The infrastructure and facilities used are supported by U.S. National Science Foundation (DMR-1106132) and the U.S. Department of Energy/National Nuclear Security Administration (DE-FC-52-08NA28554, CDAC).

\end{document}